\documentclass[twocolumn,prl,aps,superscriptaddress,showpacs]{revtex4}
\usepackage{xspace,amsmath,amsfonts,amsthm,amssymb,amsbsy,graphics,color,graphicx,epsfig}
\usepackage{epstopdf}

\newcommand{\sq}[1]{\left[ {#1} \right]}
\newcommand{\tr}[1]{{\textrm {Tr}}\sq{#1}}
\newcommand{\smallfrac}[2]{\mbox{$\frac{#1}{#2}$}}
\newcommand{\half}{\smallfrac{1}{2}}

\newcommand{\ket}[1]{|{#1}\rangle}

\newcommand{\expt}[1]{\langle{#1}\rangle}
\newcommand{\dg}{^\dagger}
\newcommand{\D}[1]{{\cal D}\sq{#1}}

\newcommand{\Hc}[1]{{\cal H}\sq{#1}}

\newcommand{\beq}{\begin{equation}} 
\newcommand{\eeq}{\end{equation}}
\newcommand{\bqa}{\begin{eqnarray}} 
\newcommand{\eqa}{\end{eqnarray}}

\newcommand{\erf}[1]{Eq.~(\ref{#1})}
\newcommand{\cu}[1]{\left\{ {#1} \right\}}

\newcommand{\an}[1]{\left\langle{#1}\right\rangle}

\begin{document}

\newtheorem{theo}{Theorem}
\newtheorem{lemma}{Lemma}

\title{Rapid Measurement of Quantum Systems using Feedback Control}

\author{Joshua Combes}
\author{Howard M. Wiseman}

\affiliation{Centre for Quantum Computer Technology, Centre for Quantum Dynamics, Griffith University, Nathan 4111, Australia} 

\author{Kurt Jacobs}

\affiliation{Department of Physics, University of Massachusetts at Boston, 100 Morrissey Blvd, Boston, MA 02125, USA}

\affiliation{Hearne Institute for Theoretical Physics, Department of Physics and Astronomy, Louisiana State University, Baton Rouge, LA 70803, USA}

\begin{abstract}
We introduce a feedback control algorithm that increases the speed at which a measurement extracts information about a $d$-dimensional system by a factor that scales as $d^2$. Generalizing this algorithm,  we apply it to a register of $n$ qubits  and show an improvement $O(n)$. We derive analytical bounds on the benefit provided by the feedback and perform simulations that confirm that this speedup is achieved. 
\end{abstract}

\maketitle 
Recently it has been shown that it is possible to increase the speed at which a measurement purifies the state of a quantum system by using real-time feedback control as the measurement proceeds~\cite{Jac0303, ComJac0601, WisRal0606, JorKor06, HilRal0705,GriHilRal07,RalGriHil06, HilRalprep,WisBou07}. Specifically, by using feedback to keep the state of the system diagonal in a basis that is unbiased with respect to the measured observable, one can make the system purity increase deterministically at a rate faster than the increase of the average purity by measurement alone~\cite{Jac0303, ComJac0601}. Since this protocol requires an unbiased basis, it exploits a purely quantum mechanical effect. However, as a consequence this quantum feedback protocol prevents the observer from obtaining full information about the initial preparation of the system, and is therefore not appropriate for use in applications such as communication channels~\cite{channel}. It is for this reason that the effect of the protocol is termed {\em rapid-purification} and not {\em rapid-measurement}. 

Here we present a new protocol that can be applied to both quantum and classical systems, and that, in contrast to quantum rapid-purification, increases the rate at which the observer gains information about the initial preparation, as well  as the rate at which the state is purified. This protocol therefore achieves not merely rapid purification, but also rapid measurement, and can be used in communication channels, state stabilisation, read out and error correction in quantum computers. It is also distinct from the rapid state-discrimination protocol introduced recently in~\cite{Jac07b}, which is only applicable to non-orthogonal states and thus to quantum systems. Our protocol can be applied to all systems with dimension $d>2$. Asumming the measured observable scales with $d$, we show that our protocol increases the speed of a measurement by a factor $O(d^2)$ over an unaided measurement on a qudit. This is in contrast to the $O(d)$ speed-up acheived by previous protocols \cite{ComJac0601}. We generalise our protocol for a register of $n$  qubits, each being measured independently and continuously, and obtain an improvement $O(n)$.

The evolution of the state $\rho$ of a system subject to a continuous measurement of an observable $X$ is given by the stochastic master equation (SME)~\cite{CMreview1,CMreview2} 
\begin{eqnarray}\label{eq1}
  d\rho & = &  
  2\gamma \, dt\, \D{X}\rho +\sqrt{2\gamma}\,dW\,\Hc{X}\rho,  
  \end{eqnarray}
where $\D{A} \rho = A\rho A\dg -\half (A\dg A \rho + \rho A\dg A)$, and  $\Hc{A} \rho =  A\rho +\rho A\dg - \tr{(A\dg+ A )\rho}\rho$. The {\em measurement strength}, $\gamma$, determines the rate at which information is extracted, and thus the rate at which the system is projected onto a single eigenstate of $X$~\cite{VanStoMab0605}. We denote the continuous measurement record obtained by the observer as $r(t)$, and $dr = \sqrt{4\gamma}\langle X \rangle dt + dW$. We assume that the measurement strength is much smaller than the strength of the Hamiltonians that can be controlled by feedback, so that applied unitaries can be treated as instantaneous.  

For a continuous measurement, without feedback, there must be no degenerate eigenvalues in the observable in order to guarantee purification~\cite{VanStoMab0605}. 
In what follows we will take $X$ to have equispaced, and hence nondegenerate, eigenvalues (although we note that, with  feedback, in order to purify an arbitrary initial state, the observable need possess only two nondegenerate eigenvalues). This is a fairly canonical choice, as it applies to observables such as the components of angular momentum and the energy of a harmonic oscillator. 
 This means that, since the transformation $X \rightarrow X + \alpha I$ for real $\alpha$ leaves the SME invariant, we can always take the $d$ eigenvalues $\cu{x}$ of $X$, to be $\{0,1,\ldots ,d-1\}$, with corresponding eigenstates $\ket{x}$. 
We will label the eigenvalues of $\rho$ as $\lambda_{j}$ in decreasing order: $\lambda_{0} \geq \lambda_{1} \geq \cdots \lambda_{d-1}$. We will assume that $\rho$ is initially completely mixed, and that any feedback merely acts so as to swap states in the eigenbasis $\cu{\ket{x}}$ of $X$. In this case $\rho$ will remain diagonal in this eigenbasis, and we will denote the eigenstate of $\rho$ with eigenvalue $\lambda_{j}$ as $\ket{x_{j}}$.

The rapid-measurement algorithms we present are based on the following two observations: 1. The rate at which a measurement distinguishes between two eigenstates, $|x\rangle$ and $|x'\rangle$ is proportional to $\gamma (x-x')^2$. 2. As the measurement proceeds, $\lambda_{0}$ approaches unity. With no feedback, the above facts mean that at long times the next-largest eigenvalue $\lambda_{1}$ will be associated with adjacent eigenvalues of $X$: $x_{1}=x_{0}\pm 1$ \cite{ComJac0601}. However, fact 1 means that if we apply a unitary operation to the state matrix so as to rearrange the eigenbasis, we can make $|x_{1}-x_{0}|$ as large as $d-1$, leading to a $(d-1)^{2}$ increase in the rate at which we distinguish between the most and next-most likely state. This is the basic idea behind our rapid measurement algorithms.

Previous analyses of rapid purification have considered both the impurity $L \equiv 1-\sum_{j}\lambda_{j}^{2}$ \cite{Jac0303,ComJac0601} and the infidelity $\Delta \equiv 1-\lambda_{0}$ \cite{WisRal0606}. In the long-time limit these are proportional ($L \sim 2\Delta \ll 1$) and for the most part we use $\Delta$ rather than $L$ in this Letter. As well 
as considering the time $\tau$ required for the {\em average infidelity} $\an{\Delta(\tau)}$ to reach a specified level $\epsilon$, one can also consider the {\em average time} $\an{T}$ for a system to attain $\Delta(T) = \epsilon$ \cite{WisRal0606}. In many circumstances the latter is the more useful quantity \cite{WisRal0606}. It is also much easier to calculate numerically. The reason is that the time $T$ at which $\Delta(T) = \epsilon$ has well-behaved statistics \cite{WisRal0606}, in contrast with $\Delta(t)$ which has extremely long tails at relatively large values of $\Delta$. That is, the average $\an{\Delta}$ is greatly influenced by the rare cases that are slow to purify. Because of this, there is substantial disagreement between $\tau$ and $\an{T}$ for a qubit. It was shown in Ref.~\cite{WisRal0606}, however, that good agreement is found between $\an{T}$ and $\tau'$, defined as the time required for $\an{\ln\Delta(\tau')}$ to reach a specified level $\ln\epsilon$. This is because taking the logarithm de-emphasizes the tails, and indeed for a qubit $ \ln\Delta$ has a near-normal distribution~\cite{WisRal0606}.

In this paper most of our calculations concern the average time $\an{T}$ to obtain a measurement fidelity $\Delta(T)=\epsilon \ll 1$. In addition to the reasons given above, this is because we can obtain  the scaling of $\expt{T}$ from analytical bounds for the long-time behaviour of $\an{\ln\Delta}$. While we are not able to obtain bounds on $\an{\Delta}$ or $\an{L}$, we include numerical simulations of $\an{L}$ for small systems. As we will see, these simulations suggest that all measures show similar increases in measurement rate under our algorithm.

We first obtain analytical expressions for $\an{\ln\Delta}$ for the case of no feedback (nfb). In the limit of interest, where one eigenvalue is very close to one, we can use the fact that $\ln\Delta\sim{\rm Tr}[\ln(1-\rho)]$. The latter expression has the advantage that one can obtain an exact integral for $\an{{\rm Tr}[\ln(1-\rho)]}$, using linear quantum trajectory theory \cite{Lintraj,CMreview2} as in Ref.~\cite{ComJac0601}. For $t\gg \gamma^{-1}$ this integral gives
\begin{eqnarray} \label{eq2} 
\expt{\ln\Delta}_{\rm nfb} \sim -4\gamma t.
\end{eqnarray}
From this relation and the argument given above (discussed in more detail in Ref.~\cite{WisRal0606}) we expect that the mean time to attain $\Delta=\epsilon$ is, for $\ln(\epsilon^{-1}) \gg 1$, 
\begin{eqnarray} \label{eq3}
\expt{T}_{\rm nfb} = (1/4\gamma)\ln(\epsilon^{-1}) .
\end{eqnarray}

We now turn to our feedback algorithm. As explained above, the basic idea is to make the eigenvalues of the measured quantity $X$ as different as possible for the eigenstates of $\rho$ with the largest and second-largest eigenvalues. Thus we take the action of the feedback unitaries to be such as to make $x_{0}$ the smallest and $x_{1}$ the largest eigenvalue of $X$. That is, $x_{0}=0$ and $x_{1}=d-1$. Then from the SME (\ref{eq1}) we   calculate  
\begin{eqnarray}\label{eq4}
d\expt{\ln\Delta}=-4\gamma dt\frac{\expt{X}^2(1-\Delta)^2}{\Delta^2},
\end{eqnarray}
where $\expt{X} = \tr{X\rho}$. For a given $\Delta$, the measurement rate is maximized by using the feedback to maximise $\expt{X}^2$. This is achieved by using feedback to order the eigenvalues of $\rho$ as $\lambda_{0},\lambda_{d-1},\lambda_{d-2},\cdots,\lambda_{2},\lambda_{1}$. Here the corresponding eigenstates are $X$-eigenstates $\ket{0},\ket{1},\ket{2}\cdots,\ket{d-2},\ket{d-1}$. We will call this L-ordering \footnote{This is based on the resemblance between the roman font capital L and the L-ordered eigenvalue profile of $\rho$.}. The feedback algorithm that maintains this ordering is Locally Optimal in time (LO) in the sense that at any point in time L-ordering will maximise the expected decrease in the log-infidelity subsequent to a weak measurent of infinitesimal duration $dt$.  It is worth noting that the feedback, applying a conditional unitary to the system, is equivalent to changing the measured observable at each time step. 

We can derive  bounds on the performance of LO feedback in the long-time limit from bounds on  
\beq
\expt{X} = \sum_{j=1}^{d-1} \lambda_{j}(d-j),
\eeq  
with the constraints $\sum_{j=1}^{d-1} \lambda_{j}=\Delta$ and $\lambda_{j}\geq \lambda_{k}\geq 0$ for $j<k$.  It is easy to prove that $\Delta (d/2) \leq \an{X}\leq \Delta (d-1)$. 
Thus in the long-time limit where $\Delta\ll 1$ we have 
\bqa \label{CLOlnD}
&&d\expt{\ln\Delta}_{\rm LO}=-4\gamma dt  S_{\rm LO}, \\
 \label{SLUB}
&&(d/2)^{2} \leq S_{\rm LO}\leq  (d-1)^{2}.
\eqa
Using this to infer approximate bounds on $\an{T}_{\rm LO}$, as explained above, we see from \erf{CLOlnD} that $S_{\rm LO}$ is precisely the speed-up-factor that this feedback algorithm gives over the measurement without feedback:
\beq \label{defSLO}
\an{T}_{\rm LO} = \an{T}_{\rm nfb} / S_{\rm LO}.
\eeq
The above argument thus predicts that the achievable speed-up in $\langle T\rangle$ increases quadratically in the size of the system. This scaling, which we confirm with numerical simulations below, is the main result of this Letter.

We expect the true speed-up factor $S_{\rm LO}$ to be closer to the lower bound in \erf{SLUB}, for the following reason. For $\Delta \ll 1$, the observer is almost certain that the system is in state $\ket{0}$, but it could be in one of the states $\ket{1},\cdots,\ket{d-1}$. For some state $\ket{x}$ with $x>0$, the larger $x$ is, the better the measurement is at determining that it is not in that state. Hence the LO feedback ensures that the measurement will tend to reduce $\lambda_{j-1}$ more than $\lambda_{j}$, $\forall j >0$. Since $\lambda_{j-1}>\lambda_{j}$, the measurement plus LO feedback will tend to equalize the $\lambda_{j}$ for $j>0$. The lower bound is attained when they are all equal. 

\begin{figure}
\leavevmode \includegraphics[width=1\hsize]{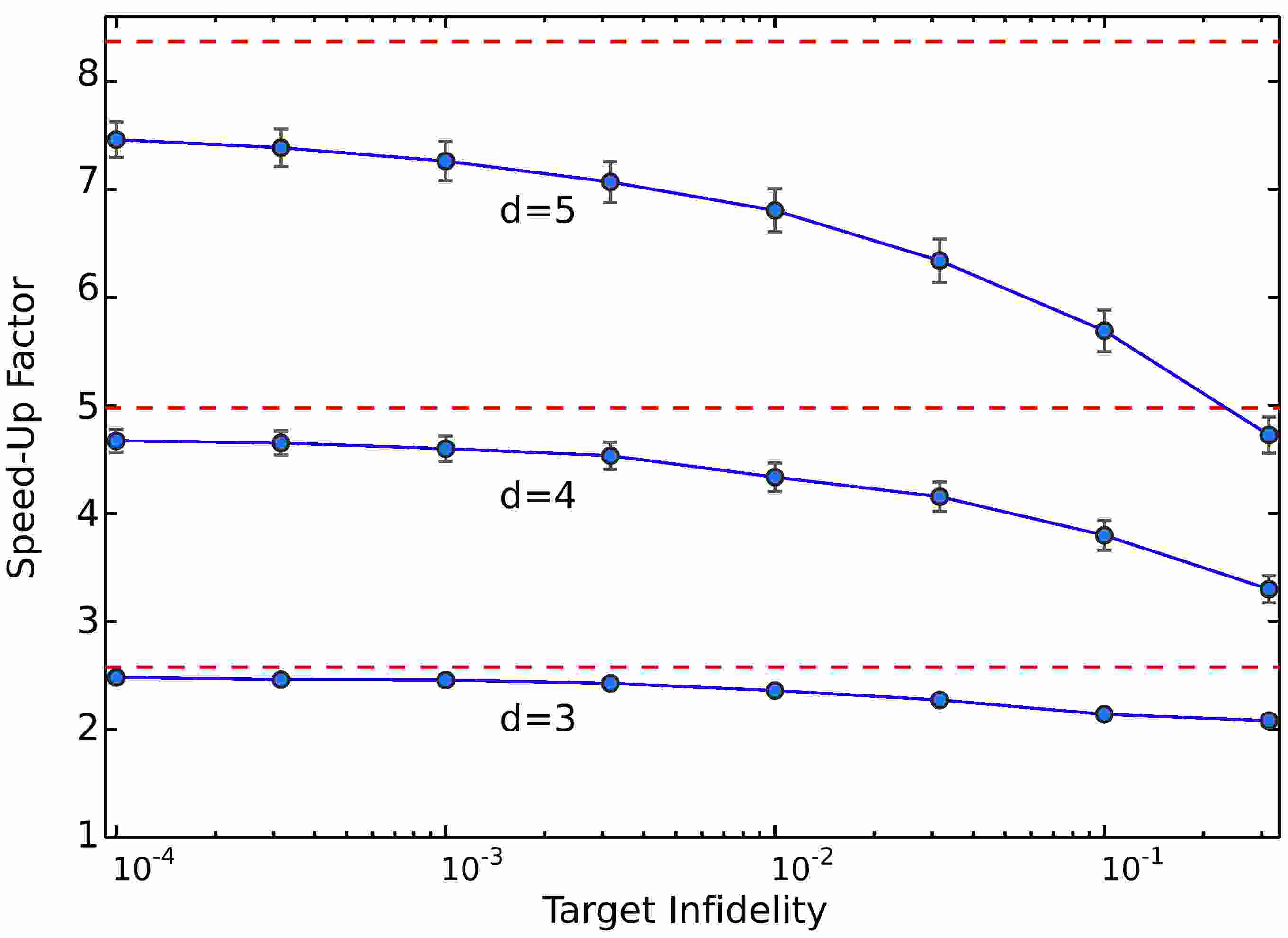}
\caption{ (color online). The speed-up in the mean time to reach a given infidelity for systems of dimension $d = 3, 4$ and $5$, when the initial state is completely mixed. The dotted lines give the numerically calculated asymptotic speedup.} \label{fig1}
\end{figure}

\begin{figure}
\leavevmode \includegraphics[width=1\hsize]{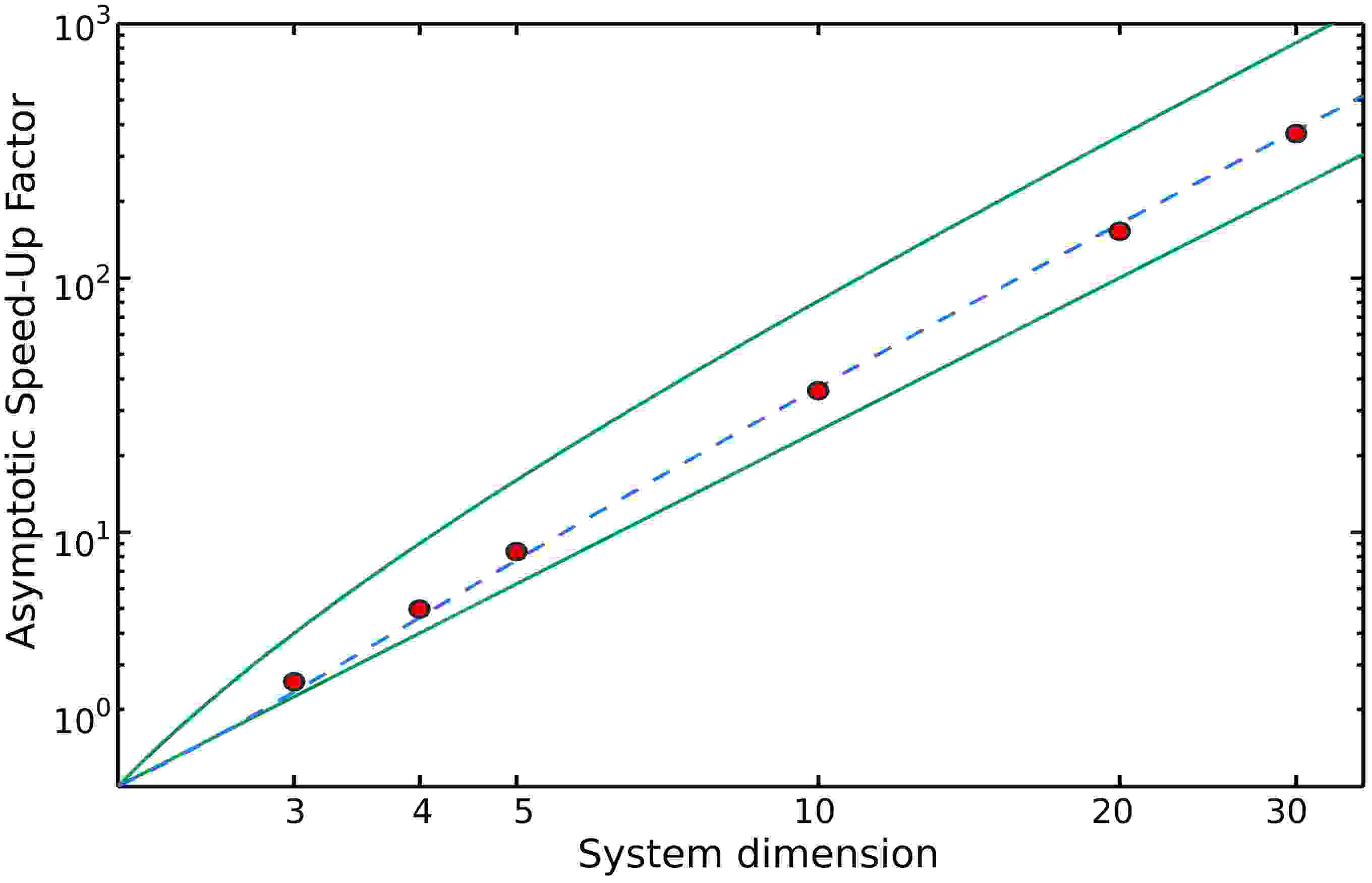}  
\caption{(color online). The asymptotic speed-up for reaching a given level of infedelity   as a function of the system dimension. The dotted line is a linear fit to the data points.   The solid lines are the upper and lower bounds derived in the text.} \label{fig2}
\end{figure}

We now compare the above calculation to numerics. We calculated $\an{T}_{\rm nfb}$ and $\an{T}_{\rm LO}$ by stochastic simulations. The former was found to agree with the theory above \erf{eq3}. In both cases, we calculated the small-$\epsilon$ asymptotic speed-up by extrapolating the numerical values of $\an{T}$.  In Fig.~\ref{fig1} we plot the speed-up factor from \erf{defSLO} as a function of $\epsilon$, for $d=3, 4, 5$, also with the asymptotic values. This shows that most of the speed-up is achieved for $\epsilon=10^{-4}$, although the gap appears to grow with $d$.  In 
Fig.~\ref{fig2} we plot the asymptotic speed-up as a function of $d$. As expected, it fits within the bounds of \erf{SLUB}, and confirms a very nearly quadratic speed-up: the fit shown is $S = 0.45 d^2-0.9d+1$.

\begin{figure}
\leavevmode \includegraphics[width=1\hsize]{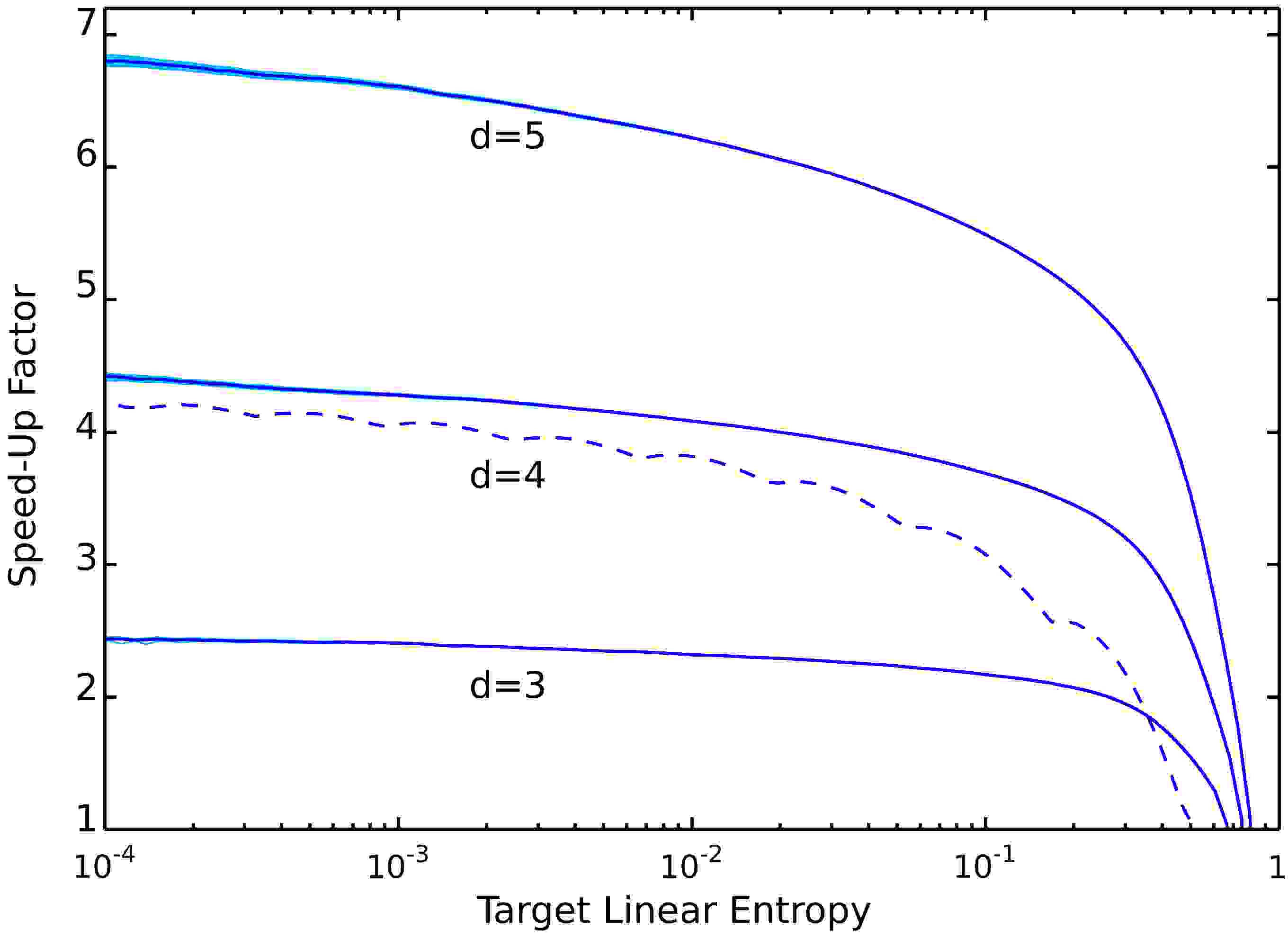}
\caption{(color online). The asymptotic speed-up in the time required to reach a given level of average purity, as a function of the final average linear entropy. Solid lines: continual feedback for systems of dimension $3,4$ and $5$; Dotted line: periodic feedback at intervals of $t=0.2/\gamma$.} \label{fig3}
\end{figure}

We also consider an alternative definition for the speed-up, based on the time taken to achieve a fixed {\em average impurity} $\expt{L}$. We calculate this speed-up factor for the LO algorithm numerically, and plot it as a function of $\langle L \rangle $ in Fig.~\ref{fig3}. While our run time is necessarily limited, these results indicate that the asymptotic speedup also falls within the bounds derived for $\langle \ln \Delta \rangle$ above. 

In an experimental implementation it may not be feasible to continually, and thus very rapidly, rearrange the elements of $\rho$. It turns out that rearranging the elements of $\rho$ at regular intervals instead is sufficient to obtain most of the speed-up. In this alternative strategy the uncertainty tends to be localized into three adjacent eigenstates by allowing the measurement to proceed {\em without} feedback for a fixed time. After the fixed time the feedback algorithm re-arranges the eigenvalues into L-ordering. We simulated this scenario for $d=4$, with re-arrangements performed at intervals of $t=0.2/\gamma$. The resulting speed-up is plotted in Fig.~\ref{fig3}, showing that the performance is not reduced significantly by this change.   We find that when the feedback is turned on the rate of decrease of $\langle L \rangle$ is in fact greater than that for the LO feedback algorithm. However, this rate decreases over time, and at no time does the alternative strategy beat the locally optimal algorithm. This suggests that the latter may be globally optimal. 
 
\begin{figure}
\leavevmode \includegraphics[width=1\hsize]{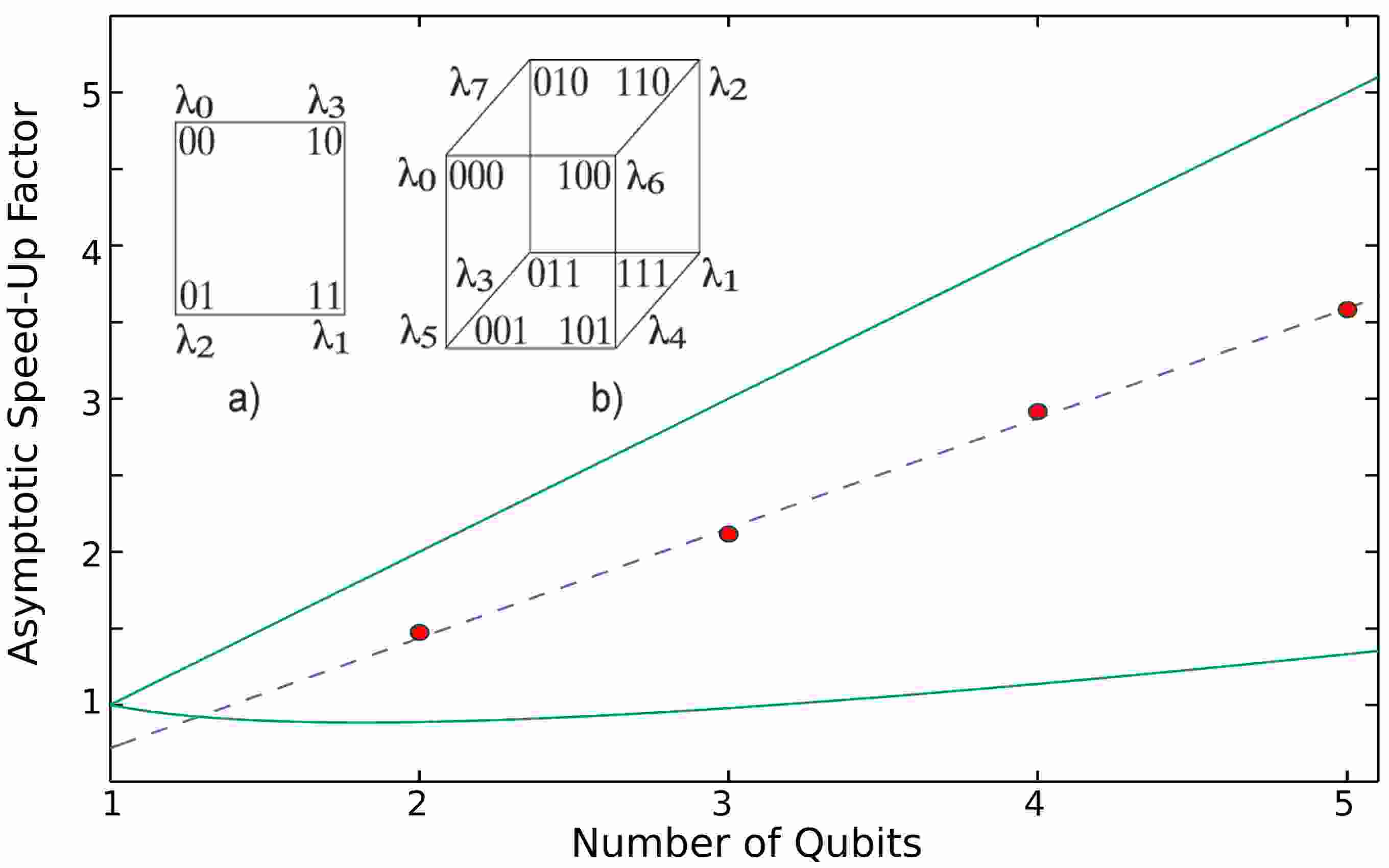}
\caption{(color online). The asymptotic speed-up in the mean time for a quantum register to reach a given level of infidelity, as a function of the number of qubits in the register.  The dashed line is a linear fit. The solid lines are the bounds derived in the text. Inset: the optimal eigenvalue arrangement for a) a two qubit register, and b)  a three qubit register.} \label{fig4}
\end{figure}

We now generalize the above rapid measurement protocol to a register of $n$ qubits, where each qubit is independently and continuously measured. Instead of one observable $X$, we now have $n$, given by $Z_r= I^{(1)}\otimes I^{(2)}\otimes \ldots \sigma_z^{(r)}\ldots \otimes I^{(n)} $, where $r$ labels the $r$th qubit. The SME describing such a measurment is \footnote{To compare with the previous results, set $\kappa = 4\gamma$.}
\begin{equation}\label{sme_reg}
  d\rho = \sum _r 2\kappa \, dt\,\D{Z_r}\rho+\sqrt{2\kappa}\,dW_r\Hc{Z_r}\rho.
\end{equation}
We transform $Z_r$ according to $Z_r\rightarrow Z_r-I$ and order the eigenvalues of $\rho$ such that $\lambda_\alpha>\lambda_\beta$ when $\alpha<\beta$, where $\alpha, \beta$ are binary strings. Without loss of generality we can assume the largest eigenvalue of $\rho$ ($\lambda_{\bar{0}}$) to be placed such that it corresponds to $\ket{\bar{0}}=\ket{00\ldots0}$. Using a similar technique to that described above \erf{eq3}, we obtain a long-time analytical expression for $\expt{\ln\Delta}$ in the absence of feedback:  $ \expt{\ln(\Delta)}_{\rm nfb}\sim -16\kappa t$. For the LO feedback algorithm, the SME (\ref{sme_reg}) is used to calculate the average rate of change of the log-infidelity, which is 
\begin{eqnarray}
d\expt{ \ln{\Delta} }&=&-4\kappa dt\frac{\sum _{r}\expt{Z_r}^2(1-\Delta)^2} {\Delta^2}.
\end{eqnarray}
Now we wish to maximize the average reduction of the log-infidelity for a given $\rho$ using feedback. This is acheived by reordering the elements of $\rho$, so as to maximize $\sum_r\expt{Z_r}^2$, in the following way.  By definition (above) the largest eigenvalue is at $\ket{\bar{0}}$. The second largest eigenvalue $\lambda_{00\ldots01}$ is then placed at $\ket{\bar{1}}$ such that it is the maximum Hamming distance \cite{Ham1950} away. The next $n$ largest eigenvalues are placed at one Hamming unit away from  $\ket{\bar{1}}$. Then the next $^nC_2$ largest eigenvalues are placed two Hamming units away from $\ket{\bar{1}}$ and so on (this process is repeated a total of $n-1$ times). We call this ordering H-ordering \footnote{Here H stands for Hamming-distance-based.}  and it is LO for the $\expt{\ln \Delta}$ measure.  Example H-orderings for a two- and three-qubit register are depicted in the  Fig.~\ref{fig4} insets.

We now bound, for a register of qubits, the amount by which the H-ordering algorithm speeds up the measurement process. To do this we must bound  $\sum_r\expt{Z_r}^2$. The upper bound corresponds to the probablity being concentrated into the eigenvalue that is maximally distant from $\ket{\bar{0}}$,  that is $\lambda_{00\ldots 1} = \Delta$, while the lower bound is reached when the probablity $\Delta$ is distributed equally over the remaining $2^n-1$ eigenvalues. The bounds are $(n2^{2n}/(2^n-1)^2) \Delta^2 \leq\sum_r\expt{Z_r}^2\leq 4n\Delta^2$. In the long-time limit ($\Delta<<1$) we thus find
\bqa 
&&d\expt{\ln\Delta}_{\rm LO}=-4\kappa dt  S_{\rm LO},\label{CLOlnDreg}\\
&& n2^{2n}/4(2^n-1)^2\leq S_{\rm LO}\leq  n . \label{SLUBreg} 
\eqa
For $n\gtrsim 7$ the lower bound is approximately $n/4$. These are rigorous bounds on the speed-up factor for $\langle \ln\Delta \rangle$, and once again we expect these to well-approximate the behavior of $\langle T \rangle$. To confirm this we performed numerical simulations of the nfb and LO algorithms for quantum registers of different sizes, and extrapolated the asymptotic speed-ups as above. The results are displayed in Fig.~\ref{fig4}, and show the expected linear dependence on the number of qubits; the linear fit is $S_{\mbox{\scriptsize H}} = 0.718 n$.

In summary, we have shown that it is possible to use feedback control during a measurement process to increase the speed of the measurement by a factor proportional to $d^2$. One can also use this method to speed up the measurement of a register of $n$ qubits by a factor that scales as $n$. We expect that approximate algorithms which are experimentally feasible should achieve most of this speedup, so that feedback-enhanced measurement could be applied in a variety of quantum technologies.

{\em Acknowledgements:} JC and HMW are supported by the Australian Research Council (FF0458313 and CE0348250)  \& the State of  Queensland.  KJ is supported by the ARO \& DTO. The authors acknowledge the use of the UMass Boston and G.U. parallel computing clusters.  

\bibliographystyle{apsrev}


\begin{thebibliography}{99}
\bibitem{Jac0303}
   K. Jacobs, Phys. Rev. A {\bf 67}, 030301(R) (2003).
\bibitem{ComJac0601}
   J. Combes and K. Jacobs, Phys. Rev. Lett. {\bf 96}, 010504 (2006). 
\bibitem{WisRal0606}
  H. M. Wiseman and J. F. Ralph, New J. Phys., {\bf 8}, 90 (2006).
\bibitem{JorKor06}
A. N. Jordan and A. N. Korotkov, Phys. Rev. B {\bf 74}, 085307 (2006).
\bibitem{HilRal0705}
C Hill and J. F. Ralph, New J. Phys., {\bf 9}, 151 (2007).
\bibitem{GriHilRal07}
E. J. Griffith et. al., Phys. Rev. B, {\bf 75}, 014511 (2007).
\bibitem{RalGriHil06}
 J. F. Ralph, E. J. Griffith, C. D. Hill and T. D. Clark, Proc. SPIE-Int. Soc. Opt. Eng., {\bf 6244}, 624403 (2006).
\bibitem{HilRalprep}
 C. Hill and J. F. Ralph, arXiv:0709.4217v1.
\bibitem{WisBou07}
    H. M. Wiseman and L. Bouten,  arXiv:0707.3001v2.
\bibitem{channel}
A. S. Holevo, IEEE Trans. Inf. Theory, {\bf 44}, 269 (1998).
\bibitem{Jac07b}
  K. Jacobs, Quant. Information Comp., {\bf 7}, 127 (2007).
\bibitem{VanStoMab0605}
  R. van Handel, J. K. Stockton and H. Mabuchi, IEEE Trans. on Automatic Control,  {\bf 50}, 768 (2005).
\bibitem{CMreview1}
   T. A. Brun, Am. J. Phys. {\bf 70}, 719 (2002). 
\bibitem{CMreview2}
   K. Jacobs and D. Steck, Contemp. Phys. {\bf 47}, 279 (2006).
\bibitem{Lintraj}
  K. Jacobs and P. L. Knight, Phys. Rev. A. {\bf 57}, 2301 (1998); H. M. Wiseman, Quantum and Semiclassical Opt. {\bf 8} , 205 (1996).
\bibitem{Ham1950}
R. W. Hamming, Bell System Technical Journal {\bf 26} 147 (1950).
\end{thebibliography}

\end{document}